# A Universal Music Translation Network

**Noam Mor, Lior Wolf, Adam Polyak, Yaniv Taigman**
Facebook AI Research

## Abstract

We present a method for translating music across musical instruments, genres, and styles. This method is based on a multi-domain wavenet autoencoder, with a shared encoder and a disentangled latent space that is trained end-to-end on waveforms. Employing a diverse training dataset and large net capacity, the domain-independent encoder allows us to translate even from musical domains that were not seen during training. The method is unsupervised and does not rely on supervision in the form of matched samples between domains or musical transcriptions. We evaluate our method on NSynth, as well as on a dataset collected from professional musicians, and achieve convincing translations, even when translating from whistling, potentially enabling the creation of instrumental music by untrained humans.

## 1 Introduction

Humans have always created music and replicated it – whether it is by singing, whistling, clapping, or, after some training, playing improvised or standard musical instruments. This ability is not unique to us, and there are many other vocal mimicking species that are able to repeat a music from hearing.

Music is also one of the first domains to be digitized and processed by modern computers and algorithms. It is therefore somewhat surprising that in the core music task of mimicry, AI is still much inferior to biological systems.

In this work we are able, for the first time as far as we know, to produce high fidelity musical translation between instruments, styles, and genres. For example[1], we convert the audio of a Mozart symphony performed by an orchestra to an audio in the style of a pianist playing Beethoven. Our ability builds upon two technologies that have recently become available: (i) the ability to synthesize high quality audio using auto regressive models, and (ii) the recent advent of methods that transform between domains in an unsupervised way.

The first technology is important for two reasons. First, it allows us to generate high quality and realistic audio. Second, trained with the teacher forcing technique, autoregressive models are efficiently trained as decoders. The second family of technologies contributes to the practicality of the solution, since posing the learning problem in the supervised setting would require a parallel dataset of different musical instruments.

In our architecture, we employ a single, universal, encoder and apply it to all inputs. In addition to the advantage of training fewer networks, this also enables us to convert from musical domains that were not heard during training to any of the domains encountered.

The key to being able to train a single encoder architecture is making sure that the domain-specific information is not encoded. We do this using a domain confusion network that provides an adversarial signal to the encoder. In addition, it is important for the encoder not to memorize the input signal but to encode it in a semantic way. We achieve this by distorting the input audio by random local pitch modulation.

---

[1]Audio samples are available at: https://ytaigman.github.io/musicnet

During training, the network is trained as a denoising autoencoder, which recovers the undistorted version of the original input. Since the distorted input is no longer in the musical domain of the output, the network learns to project out-of-domain inputs to the desired output domain. In addition, the network no longer benefits from memorizing the input signal and employs a higher-level encoding.

Our results present abilities that are, as far as we know, unheard of. Asked to convert one musical instrument to another, our network is on par or slightly worse than professional musicians. Many times, people find it hard to tell which is the original audio file and which is the output of the conversion that mimics a completely different instrument. On the encoding side, our network is able to successfully process unseen musical instruments or other sources, such as whistles. On the output side, relatively high quality audio is produced and new instruments can be added without retraining the entire network.

## 2 Previous Work

**Domain Transfer** Recently, there has been a considerable amount of work, mostly on images and text, which performs unsupervised translation between domains $\mathcal{A}$ and $\mathcal{B}$ without being shown any matching pairs, i.e., in a completely unsupervised way. Almost all of this work employs GAN constraints [1] in order to ensure a high level of indistinguishability between the translations of samples in $A$ and samples from the domain $B$. In our work, the output is generated by an autoregressive model and training takes place using the ground truth output of the previous time steps ("teacher forcing"), instead of the predicted ones. A complete autoregressive inference is only done during test time, and it is not practical to apply such inference during training in order to get a realistic generated ("fake") sample for the purpose of training the GAN.

Another popular constraint is that of circularity, namely that by mapping from $\mathcal{A}$ to $\mathcal{B}$ and back to $\mathcal{A}$ a reconstruction of the original sample is obtained [2, 3, 4]. In our work, for the same reason mentioned above, the output during training does not represent the future test time output, and such a constraint is unrealistic. An application of circularity in audio was present in [5], where a non-autoregressive model between vocoder features is used to convert between voices in an unsupervised way.

Cross domain translation is not restricted to a single pair of domains. The recent StarGAN [6] method creates multiple cycles for mapping between multiple (more than two) domains. The method employs a single generator that receives as input the source image as well as the specification of the target domain and produces the analog "fake" image from the target domain. Our work employs multiple decoders, one per domain, and attempts to condition a single decoder on the selection of the output domain failed to produce convincing results.

Another type of constraint is provided by employing a shared latent space from which samples in both domains are generated. CoGAN [7] learns a mapping from a random input vector $z$ to matching samples, one in each domain. The two domains are assumed to be similar and their generators (and GAN discriminators) share many of the layers' weights. Specifically, the earlier generator layers are shared while the top layers are domain-specific. CoGAN has applied to the task of domain translation in the following way: given a sample $x \in \mathcal{A}$, a latent vector $z_x$ is fitted to minimize the distance between the image generated by the first generator $G_{\mathcal{A}}(z_x)$ and the input image $x$. Then, the analogous image in $\mathcal{B}$ is given by $G_{\mathcal{B}}(z_x)$. Applying optimization during inference leads to slower solutions and to reliance on good initialization. On the other hand, it may lead to multiple solutions, which is sometimes desirable.

UNIT [8] employs an encoder-decoder pair per each domain, where the latent spaces of the domains are assumed to be shared. Similarly to CoGAN, the layers that are distant from the image (the top layers of the encoder and the bottom layers of the decoder) are the ones shared. Cycle-consistency is added as well, and structure is added to the latent space using a variational autoencoder [9] loss terms. Our method employs a single encoder, which eliminates the need for many of the associated constraints. In addition, we do not impose a VAE loss term[9] on the latent space of the encodings and instead employ a domain confusion loss [10].

**Audio Synthesis** WaveNet [11] is an autoregressive model that predicts the probability distribution of the next sample, given the previous samples and an input conditioning signal. Its generated output is currently considered of the highest naturalness, and is applied in a range of tasks. In [12], the authors have used it for denoising waveforms by predicting the middle ground-truth sample from its noisy input support. Recent contributions in Text-To-Speech(TTS) [13, 14] have successfully conditioned wavenet on linguistic and acoustic features to obtain state of the art performance. In our encoder-decoder architecture, we use WaveNet as the output of the decoder, and backpropagate through it down to the encoder.

In [15], voice conversion was obtained by employing a variational autoencoder that produces a quantized latent space that is conditioned on the speaker identity. Similarly to our work, the decoder is based on WaveNet [11], however we impose a greater constraint on the latent space by (a) having a universal encoder, forcing the embeddings of all domains to lie in the same space, yet (b) training a separate reconstructing decoder for each domain, provided that the (c) latent space is disentangled, thereby reducing source-target pathways memorization, which is also accomplished by (d) employing augmentation to distort the input signal.

The specific architecture of the autoencoder we employ is the wavenet-autoencoder presented in [16]. In comparison to this work, our inputs are not controlled and are collected from consumer media. Our overall architecture differs in that multiple decoders and an additional auxiliary network used for disentanglement are trained and by the introduction of a crucial augmentation step. By choosing to employ the same hyperparameters as previous work for the encoder and decoders themselves, the contribution of our approach is further emphasized.

In the supervised learning domain, an audio style transfer between source and target spectrograms was performed with sequence-to-sequence recurrent networks [17]. This method requires matching pairs of samples played on different instruments. In another fully supervised work [18], a graphical model aimed at modeling polyphonic tones of Bach was trained on notes, capturing the specificity of Bach's chorales. This model is based on recurrent networks and requires a large corpus of notes of a particular instrument produced with a music editor.

**Style Transfer** Style transfer is often confused with domain translation and many times the distinction is not clear. In the task of style transfer, the "content" remains the same between the input and the output, but the "style" is modified. Notable contributions in the field include [19, 20, 21]. These methods synthesize a new image that minimizes the content loss with respect to the content-donor sample and the style loss with respect to one or more samples of a certain style. The content loss is based on comparing the activations of a network training for an image categorization task. The style loss compares the statistics of the activations in various layers of the categorization layer. An attempt at audio style transfer is described in [22].

We distance ourselves from style transfer and do not try to employ such methods since we believe that a melody played by a piano is not similar except for audio texture differences to the same melody sung by a chorus. The mapping has to be done at a higher level and the modifications are not simple local changes.

A support to our approach is provided by the current level of success using classical conversion methods, which are still limited to monophonic instruments (one note each time). Such methods employ an analysis followed by a synthesis framework. First, the signal is analyzed to extract pitch and timbre (using harmonics tracking) and then it is converted to another monophonic instrument, using a known timbre model [23].

## 3 Method

Our method is based on training multiple autoencoder pathways, one per musical domain, such that the encoders are shared. During training, a softmax-based reconstruction loss is applied to each domain separately. The input data is randomly augmented prior to applying the encoder in order to force the network to extract high-level semantic features instead of simply memorizing the data. In addition, a domain confusion loss [10] is applied to the latent space to ensure that the encoding is not domain-specific. A diagram of the architecture is shown in Fig. 1.

### 3.1 WaveNet Autoencoder

We reuse an existing autoencoder architecture that is based on a WaveNet decoder and a WaveNet-like dilated convolution encoder [16]. The WaveNet of each decoder is conditioned on the latent representation produced by the encoder. In order for the architecture to fit the inference-time, CUDA kernels provided by NVIDIA ( https://github.com/NVIDIA/nv-wavenet) were used after slightly modifying the WaveNet equations for compatibility.

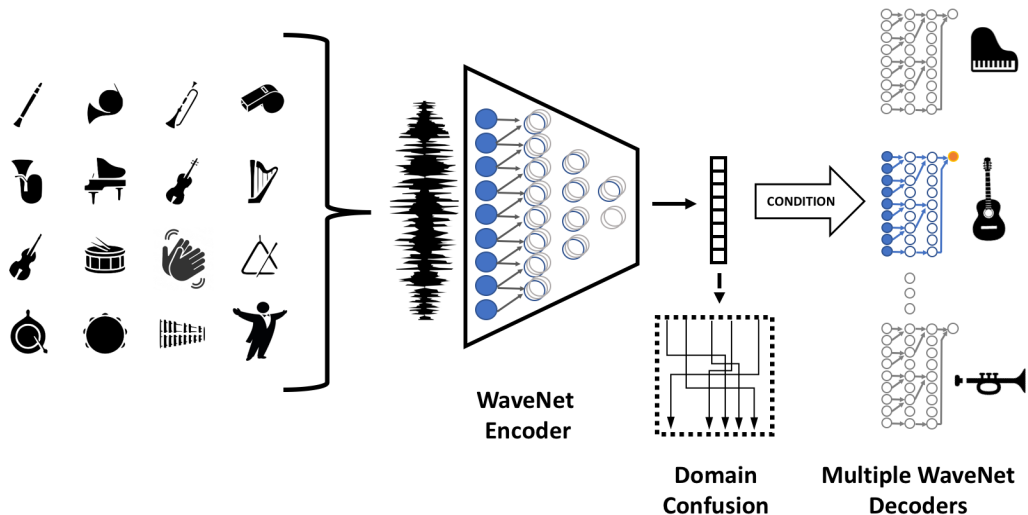

Figure 1: The architecture of our network. The confusion block (dashed line) is employed only during training.

The encoder is a fully convolutional network that can be applied to any sequence length. The network has three blocks of 10 residual-layers each. Each residual-layer contains a RELU nonlinearity, a dilated convolution with an increasing kernel size, a second RELU, and a $1 \times 1$ convolution followed by the residual summation of the activations before the first RELU. There is a fixed width of 128 channels. After the three blocks, there is an additional $1 \times 1$ layer. An average pooling with a kernel size of 50 milliseconds (800 samples) follows in order to obtain an encoding in $\mathbb{R}^{64}$, which implies a temporal down sampling by a factor of $\times 12.5$.

The encoding is upsampled temporally to the original audio rate using nearest neighbor interpolation and is used to condition a WaveNet decoder. The conditioning signal is passed through a $1 \times 1$ layer that is different for each WaveNet layer. The audio (both input and output) is quantized using 8-bit mu-law encoding, similarly to both [11, 16], which results in some inherent loss of quality. The WaveNet decoder has 4 blocks of 10 residual-layers, as a result the decoder has a receptive field of 250 milliseconds (4,093 samples).

### 3.2 Audio Input Augmentation

In order to improve the generalization capability of the encoder, as well as to enforce it to maintain higher-level information, we employ a dedicated augmentation procedure that changes the pitch locally. The resulting audio is of a similar quality but is slightly out off tune.

Specifically, we perform our training on segments of one second length. For augmentation, we uniformly select a segment of length between 0.25 and 0.5 seconds, and modulate its pitch by a random number between -0.5 and 0.5 of half-steps, using librosa [24].

### 3.3 Training and the Losses Used

Let $s^j$ be an input sample from domain $j = 1, 2, \ldots, k$, $k$ being the number of domains employed during training. Let $E$ be the shared encoder, and $D^j$ the WaveNet decoder for domain $j$. Let $C$ be the domain classification network, and $O(s, r)$ be the random augmentation procedure applied to a sample $s$ with a random seed $r$.

The $C$ network predicts which domain the input data came from, based on the latent vectors. To do so it applies three 1D-convolution layers, with the ELU [25] nonlinearity. The last layer projects the vectors to dimension $k$. The vectors are then averaged to obtain a single vector of dimension $k$.

The autoencoders $j = 1, 2, \ldots$ are trained with the loss

$$\sum_j \sum_{s^j} \mathbb{E}_r \mathcal{L}(D^j(E(O(s^j, r))), s^j) - \lambda \mathcal{L}(C(E(O(s^j, r))), j) \qquad (1)$$

where $\mathcal{L}(o, y)$ is the cross entropy loss applied to each element of the output $o$ and the corresponding element of the target $y$ separately. Note that the decoder $D^j$ is an autoregressive model that is conditioned on the output of $E$. During training, the autoregressive model is fed the target output $s^j$ from the previous time-step, instead of the generated output. The domain confusion network $C$ is trained to minimize the classification loss:

$$\sum_j \sum_{s^j} \mathbb{E}_r \mathcal{L}(C(E(O(s^j, r))), j) \qquad (2)$$

### 3.4 Network during inference

To perform the actual transformation from a sample $s$ from any domain, even from an unseen musical domain, to output domain $j$, we apply the autoencoder of domain $j$ to it, without applying the distortion. The new sample $\hat{s}^j$ is therefore given as $D^j(E(s))$. The bottleneck during inference is the autoregressive process done by the WaveNet, which is optimized by the dedicated CUDA kernels by NVIDIA.

## 4 Experiments

We describe below the training process, the datasets used for training, as well as an ablation study. Extensive experiments were done on unconstrained music as well as on the NSynth [16] dataset. Audio samples are available in the supplementary archive.

**Training** We train our network on six arbitrary classical musical domains: (i) Mozart's 46 symphonies conducted by Karl Böhm, (ii) Haydn's 27 string quartets, performed by the Amadeus Quartet, (iii) J.S Bach's cantatas for orchestra, chorus and soloists, (iv) J.S Bach's organ works, (v) Beethoven's 32 piano sonatas, performed by Daniel Barenboim, and (vi) J.S Bach's keyboard works, played on Harpsichord. The music recordings by Bach are from the Teldec 2000 Complete Bach collection. The training and test splits are strictly separated by dividing the tracks (or audio files) between the two sets. The segments used in the evaluation experiments below were not seen during training.

During training, we iterate over the training domains, such that each training batch contains 16 randomly sampled one second samples from a single domain. Each batch is first used to train the adversarial discriminator, and then to train the universal encoder and the domain decoder given the updated discriminator.

The system was implemented in the PyTorch framework, and trained on eight Tesla V100 GPUs for a total of 6 days. We used the ADAM optimization algorithm with a learning rate of $10^{-3}$ and a decay factor of 0.98 every 10,000 samples. We weighted the confusion loss with $\lambda = 10^{-2}$.

We attempted to perform two ablation studies. In the first study, the training procedure did not use the augmentation procedure of Sec. 3.2; in the second, the domain confusion network was not used ($\lambda = 0$). Both models did not train well and either diverged after some time or trained too slowly. Despite considerable effort we were not able to obtain ablation models that are compatible with further experimentation.

**Evaluation of translation quality** We consider human musicians, who are equipped by evolution, selected among their peers according to their talent, and who have trained for decades, as the gold standard and do not expect to do better than humans. To compare our method to humans, we convert from domain $X$ to piano, for various $X$. The piano was selected for practical reasons: pianists are in higher availability than other musicians and a piano is easier to produce than, e.g., an orchestra.

Three professional musicians with a diverse background were employed for the conversion task: E, who is a conservatory graduate with an extensive background in music theory and piano performance, and also specializes in transcribing music; M, who is a professional producer, composer, pianist and audio engineer who is an expert in musical transcription; and A who is a music producer, editor, and a skilled player of keyboards and other instruments.

Table 1: MOS scores (mean± SD) for the conversion tasks.

| Converter | Harpsichord→ Piano | | Orchestra→ Piano | | New domains→ Piano | |
|---|---|---|---|---|---|---|
| | Audio quality | Translation success | Audio quality | Translation success | Audio quality | Translation success |
| E | 3.89 ± 1.06 | 4.10± 0.94 | 4.02± 0.81 | 4.12± 0.97 | 4.44±0.82 | 4.13± 0.83 |
| M | 3.82 ± 1.18 | 3.75± 1.17 | 4.13± 0.89 | 4.12± 0.98 | 4.48±0.72 | 3.97± 0.88 |
| A | 3.69 ± 1.08 | 3.91± 1.16 | 4.06± 0.86 | 3.99± 1.08 | 4.53±0.79 | 3.93± 0.95 |
| Our | 2.95 ± 1.18 | 3.07± 1.30 | 2.56± 1.04 | 2.86± 1.16 | 2.36±1.17 | 3.18± 1.14 |

Table 2: Automatic quality scores for the conversion task.

| Converter | Harpsichord→ Piano | | Orchestra→ Piano | | New domains→ Piano | |
|---|---|---|---|---|---|---|
| | NCC | DTW | NCC | DTW | NCC | DTW |
| E | 0.82 | 0.98 | 0.78 | 0.97 | 0.76 | 0.97 |
| M | 0.69 | 0.96 | 0.65 | 0.95 | 0.72 | 0.95 |
| A | 0.76 | 0.97 | 0.73 | 0.95 | 0.75 | 0.94 |
| Our | 0.84 | 0.98 | 0.82 | 0.97 | 0.88 | 0.98 |

The task used for comparison was to convert 60 segments of 5 seconds each to piano. Three varied sources were used. 20 of the segments were from Bach's keyboard works, played on a Harpsichord, and 20 others were from Mozart's 46 symphonies conducted by Karl Böhm, which are orchestral works. The last group of 20 segments was a mix of three different domains that were not encountered during training – Swing Jazz, metal guitar riffs, and instrumental Chinese music. The 60 music segments were encoded by the universal encoder and decoded by the WaveNet trained on Beethoven's piano sonatas as performed by Daniel Barenboim.

In order to compare between the conversions we employed both human evaluation and an automatic score. Each score has its own limitations. The human judgment could be a mix of the assessment of the audio quality and the assessment of the translation itself. The quality of the algorithm's output is upper bounded by the neural network architecture and cannot match that of a high quality recording. The machine judgment is also limited and measures a single aspect of the conversion.

Specifically, Mean Opinion Scores (MOS) were collected using the CrowdMOS [26] package. Two questions were asked: (1) what is the quality of the audio, and (2) how well does the converted version match the original. The results are shown in Tab. 1. It shows that our audio quality is considerably lower than the results produced by humans using a keyboard connected to a computer (which should be rated as near perfect and makes any other audio quality in the MOS experiment pale in comparison). Regarding the translation success, the conversion from Harpsichord is better than the conversion from Orchestra. Surprisingly, the conversion from unseen domains is more successful than both these domains. In all three cases, our system is outperformed by the human musicians, whose conversions will soon be released to form a public benchmark.

The automatic assessment employed the pitch tracker of the librosa package [24]. For each input segment and each translation result (by a human or by the network), we extracted the pitch information. Then, we compared the input pitch to the output pitch using either the normalized cross correlation (NCC) obtained for the optimal shift, or Dynamic Time Warping (DTW) followed by a normalized correlation.

The results are presented in Tab. 2. Comparing the pitch of the output to that of the input, our method is more conservative than the human translators. The gap is diminished after the application of DTW, which may suggest that the method preserves the timing of the input in a way that humans do not.

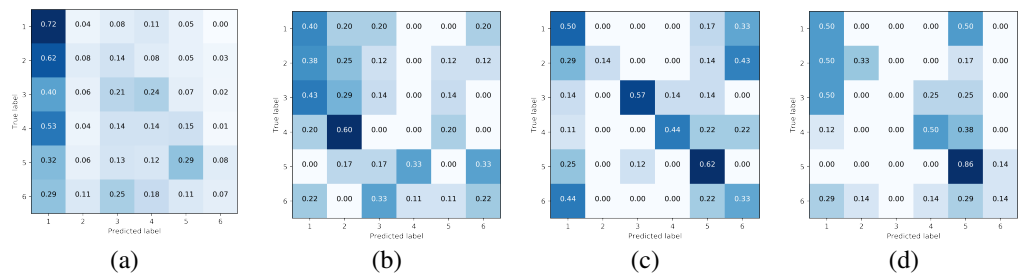

Figure 2: Results of the lineup experiment. (a) listeners from the general population tend to select the same domain as source regardless of the actual source. (b) the musician A failed to identify the source most of the time. (c) the amateurs T and (d) S failed most of the time.

**Lineup experiment**  In another set of experiments, we evaluate the ability of persons to identify the source musical segment from the conversions. We present, in each test, a set of six segments. One segment is a real segment from a random domain out of the ones used to train our network, and five are the associated translations. We shuffle the segments and ask which is the original one and which are conversions. In order to equate the quality of the source to that of the translations, we attach the source after it was passed through its domain's autoencoder.

The translation is perfectly authentic if the distribution of answers is uniform. However, the task is hard to define. In a first attempt, Amazon Mechanical Turk (AMT) freelancers tended to choose the same domain as the source regardless of the real source and the presentation order. This is shown in the confusion matrix of Fig. 2(a). We therefore asked two amateur musicians (T, a guitarist, and S a dancer and a drummer with a background in piano) and the professional musician A (from the first experiment) to identify the source sample out of the six options based on authenticity.

The results, in Fig. 2(b-d) show that there is a great amount of confusion. T and A failed in most cases, and A tended to show a similar bias to the AMT freelancers. S also failed to identify the majority of the cases, but showed coherent confusion patterns between pairs of instruments.

**Semantic blending**  The ability to blend between musical pieces in a seamless manner is one of the skills developed by DJs. It requires careful consideration of beat, harmony, volume and pitch. We use this ability in order to check the additivity of the embedding space and blend two segments linearly.

We have selected two random 5 second segments $i$ and $j$ from the Mozart symphony domain and embedded both using the encoder, obtaining $e_i$ and $e_j$. Then, we combine the embeddings as follows: starting with 3.5 second from $e_i$, we combine the next 1.5 seconds of $e_i$ with the first 1.5 seconds of $e_j$ using a linear weighting with weights $1 - t/1.5$ and $t/1.5$ respectively, where $t \in [0, 1.5]$. We then use the decoder of the Mozart symphony to generate audio. The results are natural and the shift is completely seamless, as far as we observe. See supplementary for samples.

**NSynth pitch experiments**  NSynth [16] is an audio dataset containing samples of 1,006 instruments, each sample labeled with a unique pitch, timbre, and envelope. Each sample is a four second monophonic 16kHz snippet, ranging over every pitch of a standard MIDI piano (21-108) as well as five different velocities. It was not seen during training of our system.

We measure the correlation of embeddings retrieved using the encoder of our network across pitch for multiple instruments. The first two columns (from the left hand side) of Fig. 3 show self-correlations, while the third column shows correlation across instruments. As can be seen, the embedding encodes pitch information very clearly, despite being trained on complex polyphonic audio. The cosine similarity between the two instruments for the same pitch is, on average, 0.90-0.95 (mean of the diagonal), depending on the pair of instruments.

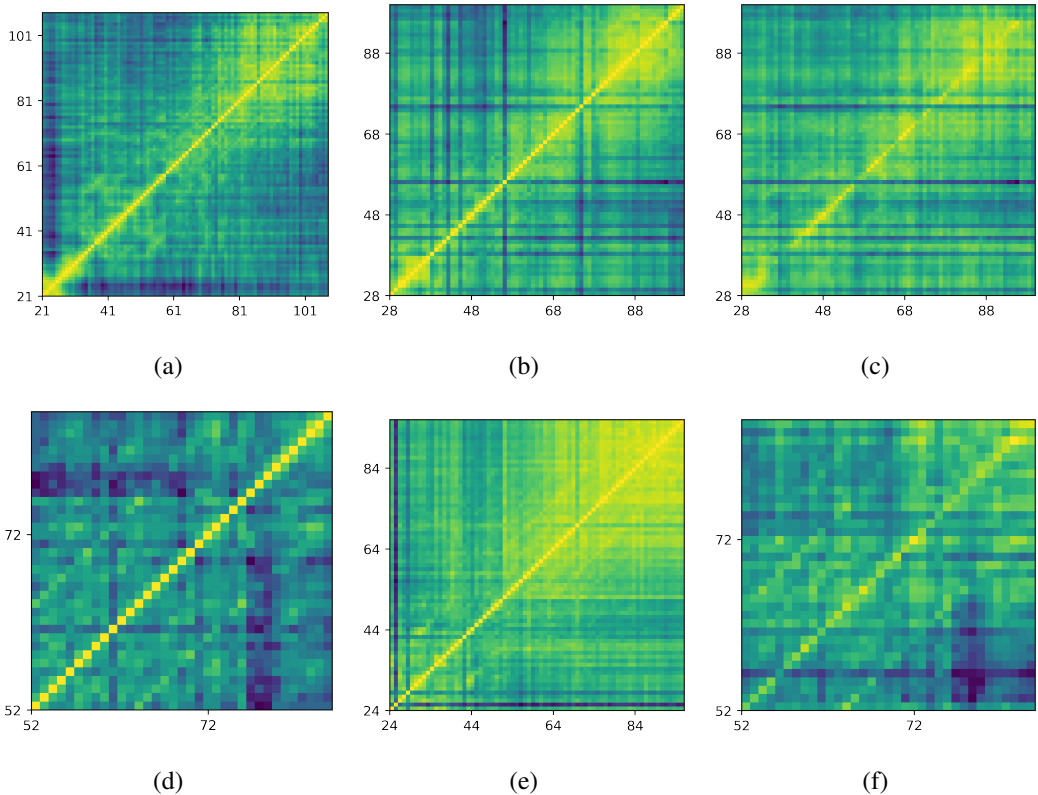

Figure 3: Correlation of embeddings across pitch. (a) Self-correlation for NSynth's flute-acoustic-027. (b) Self-correlation for keyboard-electronic-019. (c) The correlation between the electronic keyboard (y-axis) and the flute. (d) Self-correlation for brass-acoustic-018. (e) Self-correlation for string-acoustic-029. (f) The correlation between the brass instrument (y-axis) and the string.

## 5 Discussion

From a historical perspective, a universal representation has been a key component in many of the recent successes of machine learning. A notable example is AlexNet [27] and its successors, which were able to produce meaningful representations for many tasks outside ImageNet categorization. In another example, Word2Vec [28] and subsequent variants, which are trained in an unsupervised manner, are extremely effective in a wide range of NLP tasks. We are therefore encouraged by the ability of our encoder to represent, despite being trained on only six homogeneous domains, a wide variety of out-of-domain inputs.

Our work could open the way to other high-level tasks, such as transcription of music and automatic composition of music. For the first task, the universal encoder may be suitable since it captures the required information in a way, just like score sheets, that is instrument dependent. For the second task, we have initial results that we find interesting. By reducing the size of the latent space, the decoders become more "creative" and produce outputs that are natural yet novel, in the sense that the exact association with the original input is lost.

The authors of [16] have argued that while a WaveNet autoencoder cannot observe more than a fixed temporal context (around 1 second), the model is still able to produce arbitrarily long coherent audio due the continual conditioning of the encoder. We believe that these models may be able to capture additional long-term structure through the autoregressive process itself, either due to the consistency of the mapping or to being able to maintain some context. We are currently running experiments to explore this possibility.